\documentclass[11pt,twoside]{article}
\usepackage[pctex32]{graphicx}
\usepackage{epsfig}
\usepackage[figuresright]{rotating}
\usepackage{cite}
\usepackage{latexsym}
\def\tablefootnote#1{%
\hbox to \textwidth{\hss\vbox{\hsize\captionwidth\footnotesize#1}\hss}}

\usepackage{bm}
\usepackage{latexsym}
\usepackage{dcolumn}
\usepackage{amsmath,amsfonts,amssymb}
\usepackage{graphicx,epsfig}
\usepackage{psfrag}
\usepackage{amsthm}

\def\be {\begin{equation}}
\def\ee {\end{equation}}
\def\bea {\begin{eqnarray}}
\def\eea {\end{eqnarray}}
\def\bc {\begin{center}}
\def\ec {\end{center}}
\def\bfg {\begin{figure}}
\def\efg {\end{figure}}
\def\bi {\begin{itemize}}
\def\ei {\end{itemize}}
\def\nn {\nonumber}

\def\la {\label}
\def\le {\left}
\def\ri {\right}
\def\pa {\partial}
\def\fr {\frac}
\def\sq {\sqrt}

\def\a  {\alpha}

\def\d  {\delta}

\def\f {\phi}

\def\m  {\mu}
\def\n  {\nu}

\def\p  {\pi}
\def\r  {\rho}
\def\th {\theta}
\def\s {\sigma}
\def\t  {\tau}
\def\vph {\varphi}

\def\cA {\mathcal A}

\def\cH {\mathcal H}

\def\beq{\begin{equation}}
\def\eeq{\end{equation}}
\def\br{\begin{eqnarray}}
\def\er{\end{eqnarray}}
\newcommand{\eel}[1] {\label{#1}\end{equation}}

\newcommand{\bdm}{\begin{displaymath}}
\newcommand{\edm}{\end{displaymath}}

\begin{document}

\title{Where are the degrees of freedom responsible for black hole entropy?}

\author{Saurya Das$^1$\footnote{email: saurya.das@uleth.ca},~
S. Shankaranarayanan$^2$\footnote{email: shanki@aei.mpg.de} ~ and
~ Sourav Sur$^{1}$\footnote{email: sourav.sur@uleth.ca}\\ \\
{\small \em $^1$~Dept. of Physics, University of Lethbridge, 4401
University Drive, Lethbridge, Alberta, Canada T1K 3M4}\\ {\small
\em $^2$~Max-Planck-Institut f\"ur Gravitationphysik, Am
M\"uhlenberg 1, D-14476 Potsdam, Germany} }

\date{}
\maketitle

\begin{abstract}

Considering the entanglement between quantum field degrees of
freedom inside and outside the horizon as a plausible source of
black hole entropy, we address the question: {\it where are the
degrees of freedom that give rise to this entropy located?} When
the field is in ground state, the black hole area law is obeyed
and the degrees of freedom near the horizon contribute most to the
entropy. However, for excited state, or a superposition of ground
state and excited state, power-law corrections to the area law are
obtained, and more significant contributions from the degrees of
freedom far from the horizon are shown\footnote{Based on a talk by
SD at Theory Canada III, Edmonton, Alberta, Canada on June 16,
2007.}.

\end{abstract}


\section{Introduction \la{intro}}

The study of black holes (BHs) has always been a major testing
arena for models of quantum gravity. The key issue has been to
identify the microscopic origin of black hole entropy $S_{_{\rm
BH}}$. The questions that naturally arise in this context are the
following: (i) Why is $S_{_{\rm BH}}$, given by the well-known
Bekenstein-Hawking relation \cite{bek,haw},
\be \la{al1} S_{_{\rm BH}} = \le(\frac{k_{_{B}}}{4}\ri)
\frac{\cA}{\ell_P^2 } ,\qquad \le(\ell_{_P} \equiv \sqrt{{\hbar
G}/{c^3}} =  \mbox{Planck length},~ k_{_{B}} = \mbox{Boltzmann
constant}\ri) \ee
is proportional to the horizon area $\cA$, as opposed to volume
(usual for thermodynamic systems)? (ii) Are there corrections to
this so-called `area law' (AL) and if so how generic are these
corrections? (iii) Can we locate the degrees of freedom (DoF) that
are relevant for giving rise to the entropy?

In the attempts to address these questions there have been two
distinct approaches, viz., the one that associates $S_{_{\rm BH}}$
with fundamental DoF such as string, loop, etc. \cite{stringsetc},
and the other which attributes $S_{_{\rm BH}}$ to the entanglement of
quantum field DoF inside and outside the BH event horizon
\cite{bkls,sred,eisert,arom}. In this article we adopt the second
approach and consider a quantum scalar field (in a {\it pure} state)
propagating in the BH space-time. Since the BH horizon provides a
boundary to an outside observer, the state restricted outside the
horizon is {\it mixed} and leads to a non-zero entanglement ({\it aka}
Von Neumann) entropy: ~$S_{_{Ent}} = - k_{_B}$ Tr $\le(\r \ln \r\ri)$,
where $\r$ is the mixed (or {\it reduced}) density matrix obtaining by
tracing over the scalar DoF inside and outside the horizon.

In Ref. \cite{bkls,sred} --- for scalar field is in the {\it
vacuum/ground state} (GS) --- it was shown that $S_{_{Ent}}$ of
scalar fields propagating in static BH and flat space-time (the
DoF being traced inside a chosen closed surface) leads to the AL.
In Refs.  \cite{masdshanki,sdshankiES,sdshankiss}, the robustness
of the AL is examined by considering non-vacuum states. It was
shown that AL continues to hold for minimum uncertainty states
like {\it generic coherent state} (GCS) or a class of {\it
squeezed states} (SS), but for (1-particle) {\it excited states}
(ES), or for a GS-ES superposition or mixing (MS), one obtains
power-law corrections to the AL.  Although for large horizon area,
the correction term is negligible, for small BHs the correction is
significant.

To understand physically the deviation from the AL for ES/MS, we
ascertain the location of the microscopic DoF that lead to
$S_{_{Ent}}$ in these cases \cite{sdshankiss,sdshankiDoF}. We find
that although the DoF close to the horizon contribute most to the
total entropy, the contributions from the DOF that are far from the
horizon are more significant for ES/MS than for the GS. Thus, the
corrections to the AL may, in a way, be attributed to the far-away
DoF. We also extend the flat space-time analysis done in \cite{sred}
to (curved) spherically symmetric static black-hole space-times.

In the next section, we first discuss the motivation for
considering {\it scalar} fields for the entanglement entropy
computations and then show that the scalar field Hamiltonian in a
general BH space-time in Lema\^itre coordinates, and at a fixed
Lema\^itre time, reduces to that in flat space-time.  In sec.
(\ref{ent-scalar}), we briefly review the procedure of obtaining
the entanglement entropy and show the numerical results and
estimations for the cases of GS, ES and MS. In sec. (\ref{dof}),
we locate the scalar field degrees of freedom that are responsible
for the entanglement entropy and compare the results for GS and
ES/MS. We conclude with a summary and open questions in sec.
(\ref{conclu}).

In the following, we use units with $k_{_{B}} = c = \hbar = 1$ and set
$M_{_{\rm Pl}}^2 = 1/(16 \pi G)$.

\section{Hamiltonian of scalar fields in black-hole space-times}
\la{sec:BH-Ham}

Scalar fields can be motivated from the viewpoint of gravitational
perturbations in static asymptotically flat spherically symmetric
space-time background with metric $g_{\mu\nu}$.  For a metric
perturbation $h_{\mu\nu}$, the linearized form of the
Einstein-Hilbert action is invariant under the infinitesimal gauge
transformation $h_{\mu\nu} \to h_{\mu\nu} + \xi_{(\mu;\nu)}$.
Imposing the harmonic gauge condition, i.e., $\pa_{\mu} (2
h^{\mu\nu} - g^{\m\n} h_\a^\a) = 0$ \cite{barthchristen} and
keeping only the first derivatives of $h_{\mu\nu}$, one finally
obtains the linearized spin-2 equation \cite{lem}
\be
{\mathcal S}_{_{EH}}(g, h) = - \frac{M_{_{\rm Pl}}^2}{2} \int
d^4x \sqrt{|g|}\, \nabla_{\alpha} {h}_{\mu\nu} \nabla^{\alpha}
{h}^{\mu\nu} \, .
\ee
Assuming plane-wave propagation of the metric perturbations, i.e.,
$h_{\mu\nu} = M_{_{\rm Pl}} \epsilon_{\mu\nu} \varphi(x^{\mu})$
[$\epsilon_{\mu\nu} =$ polarization tensor], in the weak-field limit,
the above action reduces to the action for a massless scalar field
$\vph$ propagating in the background metric $g_{\mu\nu}$:
\be
{\mathcal S}_{_{EH}} (g, h) = - \frac 1 2 \int \!\!\!
d^4x \sqrt{|g|}\, \pa_{\alpha} \vph \pa^{\alpha} \vph \, .
\ee
Hence, by computing the entanglement entropy of the scalar fields, we
obtain the entropy of the metric perturbations of the background
space-time. The Hamiltonian of a scalar field propagating in a general
spherically symmetric space-time background with line-element:
\be \la{bh-metric}
ds^2 = - A(\tau,\xi) \, d\tau^2 + \fr{d\xi^2}{B(\tau,\xi)} +
\r^2(\tau,\xi) \le(d\theta^2 + \sin^2 \theta d\phi^2\ri) \, ,
\ee
\be
\la{eq:gen-Ham}
{\rm is~given~by} \qquad \quad
H =  \sum_{lm} \frac{1}{2} \int_{\tau}^{\infty} \!\!\!\!\!\!
d\xi \! \le[\! \frac{\sqrt{A B}}{\rho^{2}} \Pi_{_{lm}}^2 + \sqrt{A B} \, \rho^{2} (\pa_{\xi}
\vph_{_{lm}})^2 + l(l + 1) \sqrt{\frac{A}{B}} \, \vph_{_{lm}}^2 \ri] \, ,
\ee
where $A, B, \r$ are continuous, differentiable functions of
$(\tau,\xi)$ and we have decomposed $\vph$ in terms of the real
spherical harmonics $Z_{lm}(\th, \f)$, i.e., $\vph (x^{\mu}) = \sum_{l
m} \vph_{_{lm}}(\tau,\xi) Z_{l m} (\th, \f)$.

In the time-dependent Lema\^itre coordinates \cite{lem,shanki:2k3} the
line-element is given by (\ref{bh-metric}) with $A(\tau,\xi) = 1;~
B^{-1}(\tau,\xi) = 1 - f(r);~ \rho(\tau,\xi) = r(\tau,\xi)$. This
line-element is related to that in the time-independent Schwarzschild
coordinates by the following transformation relations
\cite{shanki:2k3}:
\be
\tau = t \pm \int\!\! dr \frac{\sqrt{1 - f(r)}}{f(r)} ~;~~
\xi = t + \int\!\! dr \frac{[1 - f(r)]^{-1/2}}{f(r)} ~~~~~~ \Rightarrow
\xi - \tau = \int \frac{dr}{\sqrt{1 - f(r)}}
\label{eq:xitau}
\ee
As opposed to the Schwarzschild coordinate, the Lema\^itre
coordinate is not singular at the horizon $r_h$, and $\xi$(or,
$\tau$) is space(or, time)-like everywhere while $r$(or, $t$) is
space(or, time)-like only for $r > r_h$. Choosing a fixed
Lema\^itre time ($\tau = \tau_0 = 0$), the relations
(\ref{eq:xitau}) lead to: ~$d\xi/dr = 1/\sqrt{1 - f(r)}$. Plugging
this in Eq.(\ref{eq:gen-Ham}) and on performing the canonical
transformations: ~$\Pi_{_{lm}} \to {r \sqrt{1 - f(r)}} \,
\Pi_{_{lm}} \, ; \, \vph_{_{lm}} \to \vph_{_{lm}}/r$, the
Hamiltonian reduces to that of a free scalar field propagating in
flat space-time \cite{melnikov}
\be
H = \sum_{lm} \fr 1 2 \int_0^\infty dr
\le\{\p_{lm}^2(r) + r^2 \le[\fr{\pa}{\pa r} \le(\fr{\varphi_{lm}
(r)}{r}\ri)\ri]^2 + \fr{l(l+1)}{r^2}~\varphi_{lm}^2(r)\ri\}.
\la{ham2}
\ee
This holds for {\it any} fixed $\t$, provided the scalar field is
traced over either the region $r \in (0, r_h]$ or the region $r \in
[r_h, \infty)$. Hence, evaluating the entanglement entropy of the
scalar field in flat space-time corresponds to the evaluation of
entropy of BH perturbations at a fixed Lema\^itre time.

\section{Entanglement entropy of scalar fields}
\la{ent-scalar}

We discretize the scalar field Hamiltonian (\ref{ham2}) on a radial
lattice with spacing $a$:
\bea \label{disc1}
H = \sum_{lm} H_{lm} = \sum_{lm} \fr 1 {2a} \sum_{j=1}^N \le[ \p_{lm,j}^2 +
\le(j + \fr 1 2\ri)^2 \le(\fr{\varphi_{lm,j}}{j} -
\fr{\varphi_{lm,j+1}}{j+1}\ri)^2 +
\fr{l(l+1)}{j^2}~\varphi_{lm,j}^2
\ri]  \quad ,
\eea
where $\p_{lm,j}$ are the momenta conjugate of $\vph_{lm,j}$, $(N+1)
a$ is the infrared cut-off.  $H_{lm}$ in Eq.(\ref{disc1}) is of the
form of the Hamiltonian of $N-$coupled harmonic oscillators (HOs):
\be \la{coupledham1}
H ~=~ \fr 1 2 \sum_{i=1}^N p_i^2
~+~ \fr 1 2 \sum_{i,j=1}^N x_i K_{ij} x_j \, ,
\ee
where the interaction matrix $K_{ij}$ is given by:
\bea \la{kij}
K_{ij} &=&  \fr 1 {i^2} \le[l(l+1)~\delta_{ij} + \fr 9 4 ~\d_{i1}
\d_{j1} + \le( N - \fr 1 2\ri)^2 \d_{iN} \d_{jN} +
2 \le(i^2 + \fr 1 4\ri)
\d_{i,j(i\neq 1,N)}\ri] \nn\\
&&- \le[\fr{(j + \fr 1 2)^2}{j(j+1)}\ri] \delta_{i,j+1} -
\le[\fr{(i + \fr 1 2)^2}{i(i+1)}\ri] \delta_{i,j-1} .
\eea
The last two terms denote nearest-neighbour interactions
and originate from the derivative term in (\ref{ham2}).

The most general eigenstate of the Hamiltonian (\ref{coupledham1}) is a
product of $N-$HO wave functions:
\be \la{excwavefn1}
\psi (x_1,\dots,x_N) = \prod_{i=1}^N \fr{k_{Di}^{1/4}}{\p^{1/4}~\sqrt{2^{\n_i}
\n_i!}} ~\cH_{\n_i} \le(k_{Di}^{1/4}~{\underbar x}_i \ri)
\exp\le( -\fr 1 2 k_{Di}^{1/2}~{\underbar x}_i^2 \ri),
\ee
where ${\underbar x} = Ux$, ($U^TU=I_N$), $K_D \equiv U K U^T$
(diagonal), and $\nu_i \, (i=1 \dots N)$ are the indices of the
Hermite polynomials ($\cH_{\nu}$).  The frequencies are ordered such
that $k_{Di} > k_{Dj}$ for $i > j$.

The reduced density matrix is obtained by tracing over first $n$ of the $N$
oscillators:
\bea \la{denmatgen1}
\r \le(x; x'\ri) = \int \prod_{i=1}^n dx_i ~ \psi(x_1,\dots,x_n; x_{n+1},\dots,x_N)
~\psi^\star(x_1,\dots,x_n; x'_{n+1},\dots,x'_N) ~.
\eea
It is not possible to obtain a closed form expression for $\r (x; x')$
for an arbitrary state (\ref{excwavefn1}). We resort to the following
cases to compute the entropy numerically\footnote{The computations are
done with a precision of $0.01\%$ and for $N = 300$ and $n = 100 -
200$.} using the relation $S = $ Tr $\le(\r \ln \r\ri)$:
\begin{description}
\item (i) Ground state (GS) with $N$-particle wave-function: ~$\psi_0
(x;x') \sim \exp\le[- \fr 1 2 \sum_{i=1}^N k_{Di}^{1/2} {\underbar
x}_i^2\ri]$.
\item (ii) Excited (1-particle) state (ES) with $N$-particle
wave-function: ~$\psi_1 (x;x') = \sq{2} \a^T K_D^{1/2} {\underbar x}~
\psi_0 (x;x')$, where $a^T = \le( a_1,\dots, a_N \ri)$ are the
expansion coefficients and normalization of $\psi_1$ requires $a^T
a=1$). We choose $a^{T} = 1/\sqrt{o}(0, \cdots 0, 1 \cdots 1)$ with
the last $o$ columns being non-zero.
\item (iii) GS-ES linearly superposed (i.e. mixed) state (MS) with
$N$-particle wave-function: ~$\psi (x;x') = c_0 \psi_0 (x;x') + c_1
\psi_1 (x;x')$. Normalization of $\psi$ requires constants $c_0$ and
$c_1$ related by $c_0^2 + c_1^2 = 1$. For simplicity, we choose $c_0 =
1/2$. (See details in Ref. \cite{sdshankiss}).
\end{description}
%
\begin{figure*}[!htb]
\begin{center}
\epsfxsize 6 in
\epsfysize 1.7 in
\epsfbox{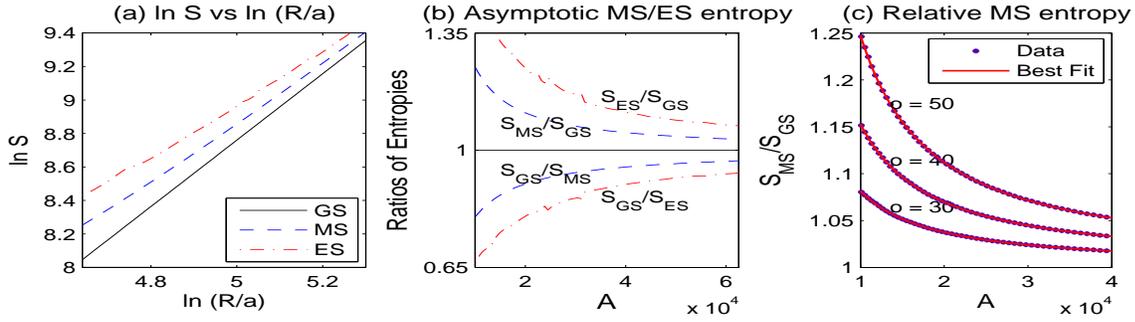}
\caption{(a) log(entropy) vs $\log(R/a)$ for GS, MS and ES
(Eq/Hi). $R = a (n + 1/2)$ is the horizon radius and $N = 300, n =
100 - 200, o = 50$. (b) Plots of $S_{MS}/S_{GS}, S_{ES}/S_{GS}$
and $S_{GS}/S_{MS}, S_{GS}/S_{ES}$ (for $o = 50$) with $\cA$ to
show the asymptotic nature of MS and ES entropies. (c) Best fit
plots of $S_{MS}/S_{GS}$ vs $\cA$ for $o = 30, 40, 50$. }
\label{fig:1}
\end{center}
\vspace*{-0.05cm}
\end{figure*}
\noindent
\underline{\it Results :} ~For GS, one recovers the AL --- $S_{GS}
\sim \cA/a^2$, where $a$ is the ultraviolet cutoff at the horizon (set
to be $\simeq \ell_P$). For MS and ES, we obtain power-law corrections
to the AL:
\be \la{p-law}
S_{MS/ES} = S_{GS} + \s \le(\cA/a^2\ri)^{1-\n} ~,
\ee
where $\s = $ constant of order unity and $\n$ is a fractional index
which depends on the excitation $o$. As the horizon area $\cA$
increases, the correction term becomes negligible and $S_{MS}
\rightarrow S_{GS}$ asymptotically.  For small BHs, however, the
correction is significant. Fig. \ref{fig:1} shows the logarithm of
entropy versus log$(R/a)$ characteristics for GS, MS and ES ($R$ being
the horizon radius), as well as the asymptotic equivalence of GS and
MS/ES entropies, and the numerical fit that leads to the above result
(\ref{p-law}).

\section{Location of the degrees of freedom}
\la{dof}

Let us take a closer look at the interaction matrix $K_{ij}$,
Eq.(\ref{kij}), for the system of $N$ HOs. The last two terms which
signify the nearest-neighbour (NN) interaction between the
oscillators, are solely responsible for the entanglement entropy of
black holes. Let us perform the following operations:
%
\begin{figure}[!htb]
\begin{center}
\epsfxsize 6 in
\epsfysize 1.3 in
\epsfbox{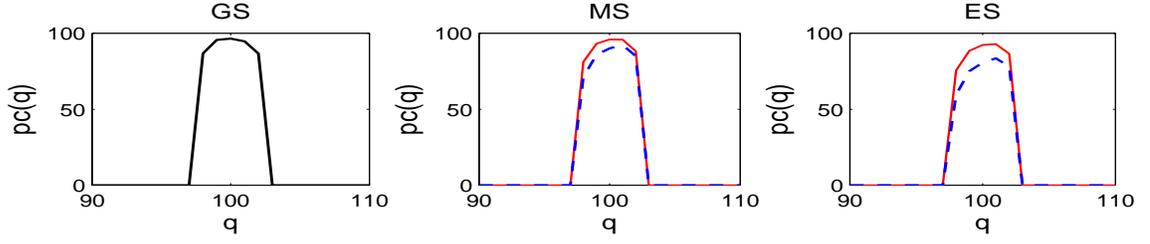}
\caption{Plots of the percentage contribution $pc(q)$ to the total
entropy as a function of window position $q$, for a window size $t
= 5$ and fixed $N = 300, n = 100$, in each of cases of GS, MS and
ES. For MS and ES the solid curve is for $o = 30$ whereas the
broken curve is for $o = 50$.}
\label{fig:2}
\end{center}
\vspace*{-0.55cm}
\end{figure}
%
\begin{figure}
\includegraphics[width=3in,height=2.5in]{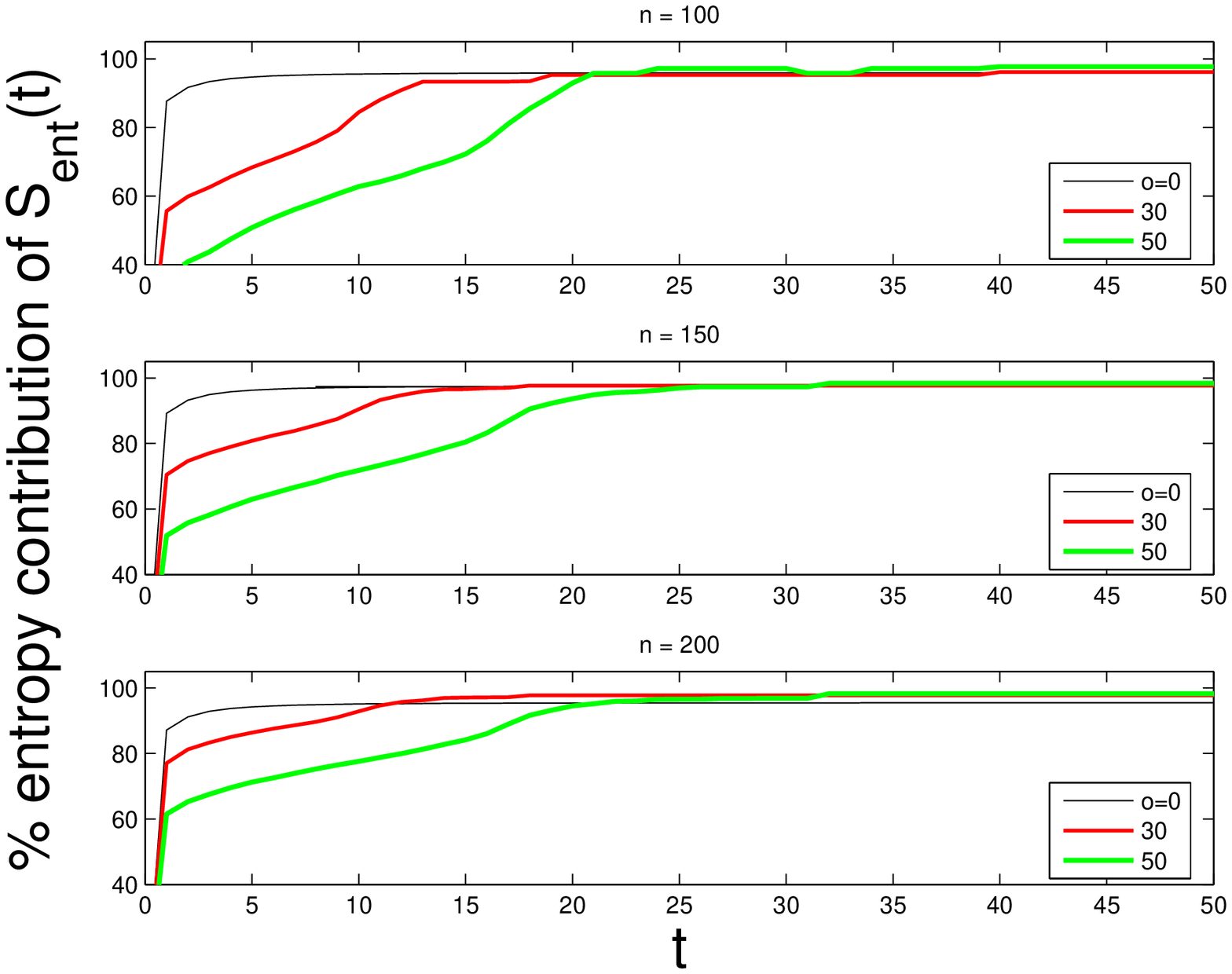}
\includegraphics[width=3in,height=2.5in]{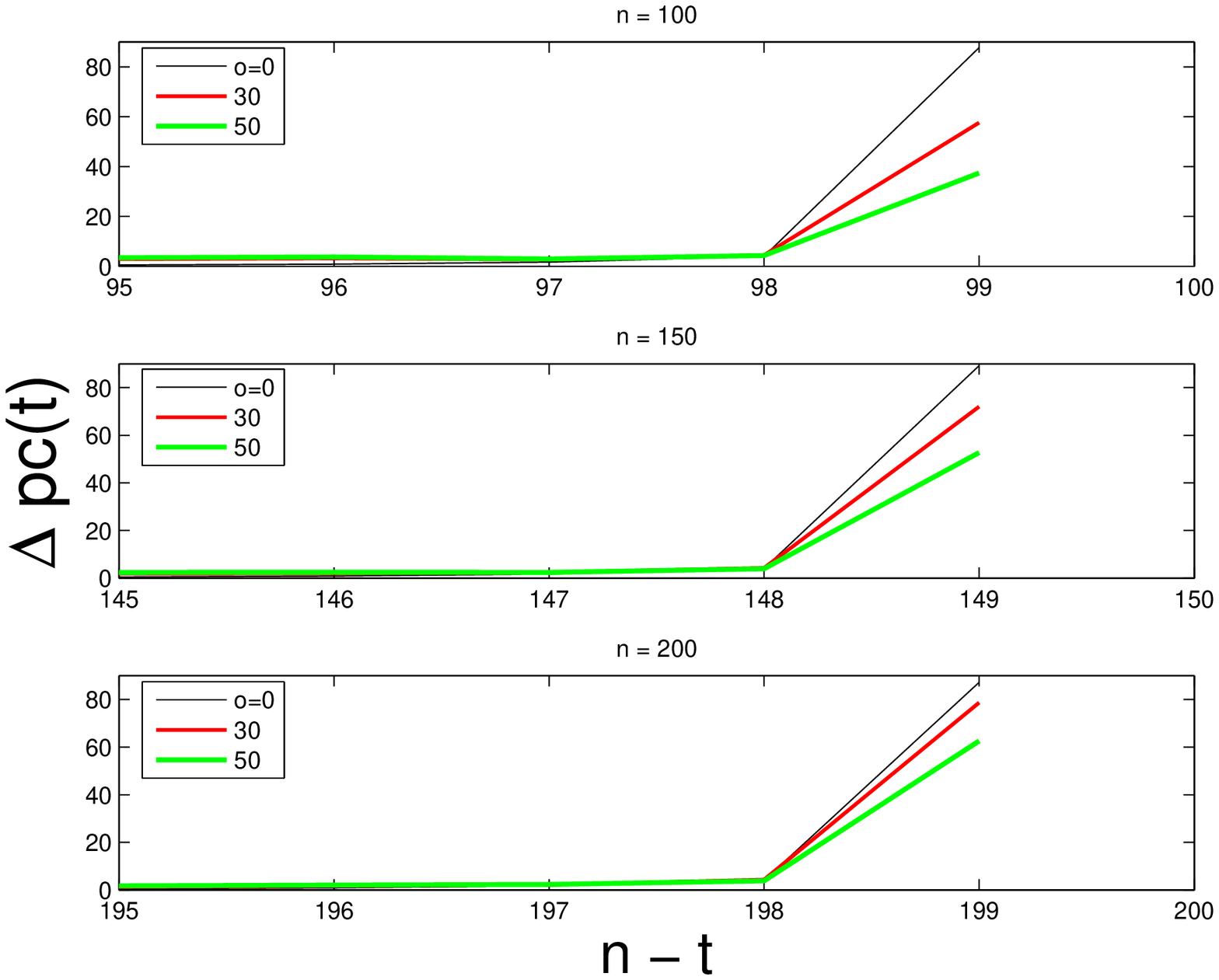}
\caption{The left panels show the variations of the percentage
contribution $pc(t)$ of total entropy with window width $t$ for GS ($o
= 0$) and ES (with $o = 30, 50$).  The right panels show the plots of
$\Delta pc(t)$ vs $n- t$ for GS and ES. Both sets of panels are for
$N=300$ and $n = 100, 150, 200$.}
\label{fig:3}
\end{figure}
\underline{\it Operation I :}~ We set NN interactions to zero (by
hand) everywhere except in a `window', such that the indices $i,j$
run from $q - s$ to $q + s$, where $s \leq q$. We thus restrict
the thickness of the interaction region to $t = 2s + 1$ radial
lattice points, while allowing it to move rigidly across from the
origin to a point outside the horizon. The variation of the
percentage contribution of the total entropy $S_{tot}$ for a fixed
window size of $t = 5$ lattice points, i.e., $pc (q) = [S (q,
t=5)/S_{tot}] \times 100$, as a function of $q$ is shown in Fig.
\ref{fig:2} for $N = 300, n = 100$, in each of the cases GS and
MS, ES with $o = 30, 50$.  In all the cases $pc (q) = 0$ when $q$
is far away from $n$ (i.e., horizon), whereas for values of $q$
very close to $n$ there are significant contributions to
$S_{tot}$. For GS, $pc (q)$ peaks exactly at $q = n$. For MS and
ES, however, the peaks shift towards a value $q > n$, and the
amplitudes of the peaks also decrease as the amounts of excitation
$o$ increase \cite{sdshankiDoF,sdshankiss}.  Therefore (a) the
near-horizon DoF contributes most to $S_{tot}$, and (b) the
contributions from the far away DoF are more for MS and ES, than
for GS.

\underline{\it Operation II :}~ We set the NN interactions to zero (by
hand) everywhere except in a window whose center is fixed at $p \leq
i,j \leq n$, and the window thickness $t\equiv n-p$ is varied from $0$
to $n$, i.e., from the origin to the horizon. For GS we find that
about $85\%$ of the total entropy is obtained within a width of just
one lattice spacing, and within a width of $t = 3$ the entire GS
entropy is recovered. Thus most of the GS entropy comes from the DoF
very close to the horizon and a small part has its origin deeper
inside. For ES, however, the total entropy is recovered for much
higher values of $t$ (than for GS) since the DoF that are away from
the horizon contribute more and more as the excitation $o$
increases. Thus larger deviation from the area law may be attributed
to larger contribution to the total entropy from the DoF far from the
horizon. The left panels of Fig. \ref{fig:3} depict the variation of
the percentage contribution to the total entropy, i.e., $pc(t) = [S
(t)/S_{tot}] \times 100$, as a function of $t$ for GS ($o = 0$) and ES
(with $o = 30, 50$). The situation is intermediate for MS (which
interpolates between the GS and the ES), i.e., the total entropy is
recovered for values of $t$ greater than that for GS, but less than
that for ES (with same value of $o$). The percentage increase in
entropy when the interaction region is incremented by one radial
lattice point, $\Delta pc(t) = pc(t) - pc(t-1)$, versus $(n-t)$ plots
for GS and ES are shown in the right panels of Fig. \ref{fig:3}.  In
the case of GS the inclusion of the first lattice point just inside
the horizon leads to an increase from $0$ to $85\%$ of the total GS
entropy.  The next immediate points add more to this, but the
contributions are lesser and lesser with inclusion of points further
and further from the horizon. For ES however, inclusion of one lattice
point adds $70(50)\%$, for $o=30(50)$, to the entropy, while the next
immediate points contribute more than those for the GS.

\section{Conclusions}
\la{conclu}

We have thus shown that if the black hole entropy is looked upon
as that due to the entanglement between scalar field DoF inside
and outside the horizon, there are power-law corrections to the
Bekenstein-Hawking area law when the field is in excited state or
in a superposition of ground state and excited state. Although
such corrections are negligible for semi-classical BHs, they
become increasingly significant with decrease in horizon area as
well as for increasing excitations. The deviation from the AL for
ES and MS may be attributed to the fact that the scalar field DoF
that are farther from the horizon contribute more to the total
entropy in the ES/MS cases than in the case of GS. The near
horizon DoF contribute most in any case, however. We have also
extended the flat space-time analysis done in \cite{sred} to
static spherically symmetric black hole space-times with
non-degenerate horizons.

We conclude with some open questions related to our work: (i) Can a
temperature emerge in the entanglement entropy scenario and would it
be consistent with the first law of BH thermodynamics? (ii) Is $dS/dt
\geq 0$?, i.e., the second law of thermodynamics valid?  (iii) Will
the entanglement of scalar fields help us to understand the
information loss problem?  We hope to report on these in future.

\bigskip
\noindent
{\large \bf Acknowledgments}

SD and SSu are supported by the Natural Sciences and Engineering
Research Council of Canada. SD thanks the organizers of Theory
Canada III, Edmonton, AB, Canada for hospitality.

\end{document}